\documentclass[aps,prl,twocolumn,showpacs]{revtex4}
\usepackage{graphics,color}
\usepackage{amsmath,amssymb}
\usepackage{wasysym,stmaryrd}

\newcommand{\dr}{\textrm{d}}
\newcommand{\ir}{\textrm{i}}

\renewcommand{\Im}{\textrm{Im}\,}
\renewcommand{\Re}{\textrm{Re}\,}

\begin{document}
\today
\title{Boundary losses and spatial statistics of complex modes in a chaotic microwave cavity}
\author{J\'er\^ome Barth\'elemy, Olivier Legrand, Fabrice Mortessagne}
\affiliation{Laboratoire de Physique de la Mati\`ere Condens\'ee,
CNRS UMR 6622,\\  Universit\'e de Nice-Sophia Antipolis, 06108 Nice cedex 2,
France}

\begin{abstract}
We experimentally study the various manifestations of ohmic losses in a
two-dimensional microwave chaotic cavity and exhibit two different
contributions to the resonance widths. We show that the parts of
these widths, which vary from mode to mode, are associated to ohmic losses located at the boundary of the cavity. We also describe how this \emph{non-proportional} damping is responsible for the complex character of wavefunctions (corresponding to a spatially non-uniform phase), which is ubiquitous in open or dissipative wave
systems. We experimentally demonstrate that the \emph{non-proportional} widths are related to a single parameter, which measures the amount of complexity of
wavefunctions, and provide theoretical arguments in favor of this relation.
\end{abstract}

\maketitle

Investigations of the morphology of wavefunction in partially open chaotic wave systems have recently become an active domain of research, especially through theoretical works concerned with quantum transport in the ballistic regime\cite{Ishio01, Seba97}. That the spatial properties of resonances are closely related to their spectral widths has been recognized ever since the pioneering work of Porter and Thomas\cite{PT56} until more recent experimental investigations in a superconducting microwave cavity\cite{Alt95}. These experiments resort to a chi-squared distribution of widths based on results of Random Matrix Theory (RMT), which is known to describe spectral and spatial statistics of closed chaotic wave systems\cite{Dittes00}.
Microwave cavities turn out to be ideal model systems to study open wave systems\cite{Stoeckbook,Barth_2002} since they offer the possibility of opening channels through antennas or connected waveguides. They can also be qualified as open because of dissipation through ohmic losses at their boundaries. Locally, boundaries may be viewed as distributed equivalent opening channels with ad hoc impedances. Through such losses, the standing waves of a closed cavity are changed into a mixture of standing and traveling waves, the latter being associated to the energy flowing out of the system\cite{PniniShap96}. \emph{Proportional} damping, \textit{i.e.} the same decay rate for all modes in a given frequency band, can only lead to modes that are real (up to a global phase) and associated to the standing wave part of the wavefunction. Hence the observation of complex-valued wavefunctions  implies \emph{non-proportional} damping mechanisms giving birth to the traveling part of the wavefunction\footnote{Note that the effect we study here is not related to the crossover of
wavefunction statistics from Gaussian Orthogonal to Unitary Ensemble of RMT
implied by the breaking of time-reversal symmetry recently investigated in
microwave experiments\cite{Chung00}.}. A thorough  description of the link between \emph{non-proportional} damping and  the complex character of wavefunctions is still lacking, as recently emphasized by Lobkis and Weaver in the context of acoustical reverberant dissipative system\cite{Lobkis_2000}.

In this letter, we present experiments in two-di\-men\-sional (2D) chaotic microwave cavities at room temperature  where dissipation is essentially due to ohmic losses at boundaries. In such chaotic cavities, modes are generically ergodic, \textit{i.e.} present speckle-like intensity patterns.  The importance of the imaginary part of the modes implied by the presence of losses can be measured through a single statistical parameter, namely the ratio $\langle(\Im\psi)^2\rangle/\langle(\Re\psi)^2\rangle\equiv q^2$ where the wavefunction $\psi$ is conventionally assumed to be real ($q=0$) in the case of vanishing losses. The principal aim of this paper is to demonstrate the intimate relation between this parameter $q$ and the part of the decay rate (or modal width) associated to \emph{non-proportional} damping. We provide strong experimental evidences of this relationship by carefully analyzing hundreds of resonances in a Sinai-like chaotic cavity. 

The cavity is the slice of a rectangular waveguide closed at both ends, with contour $\mathcal{C}$ of length $L=2.446\,$m, section $\mathcal{S}$ of area $A=0.352762\,$m$^2$ and thickness $d=5\,$mm. Inside this \emph{flat} parallelepipedic cavity is inserted an obstacle with the shape of a disk of radius 6\,cm and thickness $d$. As long as the wavelength $\lambda=2\pi k^{-1}$ is larger than $d$, the cavity may be considered as two-dimensional (2D) and only admits transverse magnetic modes. If the conductor is perfect, the electric field $\vec{E}= E^{(0)}(x,y)\hat{z}$ is perpendicular to the plane $(x,y)$ of the cavity, the magnetic field is parallel and there is no dissipated power. This ideal situation will correspond to the zeroth order of our description of the field near the surface of the actual conductor. Denoting $\psi_0 =  E^{(0)}(x,y)$, the time-independent Maxwell equations are reduced to a 2D Helmholtz equation:
\begin{equation}
\label{helm2d}
(\vec{\nabla}^2 + \epsilon\mu\omega^2)\psi_0 = 0\,,
\end{equation}
where $\epsilon$ and $\mu$ are respectively the permittivity and the permeability inside the cavity. On the contour $\mathcal{C}$, $\psi_0$ obeys Dirichlet conditions $\psi_0 = 0$.

Ohmic losses at the contour are easily computed by following a boundary perturbation technique\cite{jackson}. Indeed, the perturbed boundary conditions (BC) now read: 
\begin{equation}
\label{mixedbc}
\psi \simeq -(1+\ir) \frac{\mu_c \delta}{2 \mu} \partial_n\psi \ \textrm{on}\  \mathcal{C}\,,
\end{equation}
where $\delta = \sqrt{2/\mu_c\sigma_{cont}\omega}$ is the skin depth ($\mu_c$ being the magnetic permeability of the conductor and $\sigma_{cont}$ the effective conductivity of the contour), and where $\partial_n$ denotes $\hat{n} \cdot \vec{\nabla}$ ($\hat{n}$ being the unit normal vector directed toward the interior of the conductor). Green's theorem applied to this 2D problem straightforwardly  yields:
\begin{equation}
\label{eigenpert}
\omega^2 - \omega_0^2  \simeq - (1+i) \frac{1}{\epsilon\mu} \frac{\mu_c \delta}{2 \mu} \frac{\oint_\mathcal{C}\dr \ell\,|\partial_n\psi |^2}{\iint_\mathcal{S}\dr a\,|\psi |^2} \,,
\end{equation}
where  $\omega$ is the perturbed eigenfrequency, $\omega_0$ being the corresponding real eigenfrequency of the lossless problem. Writing  $\omega = \omega_0 + \delta\omega - i\Gamma/2$, equation (\ref{eigenpert}) becomes:
\begin{equation}
\label{eigenpert2}
\delta\omega - i\Gamma/2  \simeq - (1+i) \frac{1}{\epsilon\mu} \frac{\mu_c \delta}{4 \mu \omega_0} \frac{\oint_\mathcal{C}\dr \ell\,|\partial_n\psi |^2}{\iint_\mathcal{S}\dr a\,|\psi |^2} \,,
\end{equation}
leading, in the case at hand, to equal corrections on both real and imaginary parts of $\omega$. In our context $\mu_c/\mu$ is practically equal to unity. A similar perturbative approach can be applied to the ohmic losses at the top and the bottom of the cavity. One finds that they lead to homogeneous \emph{proportional} damping. The total width due to ohmic losses $\Gamma^\Omega$ can thus be split into two parts\cite{bigone}:
\begin{equation}
\label{gamma_ohm}
\begin{split}
\Gamma^{\Omega} &= \Gamma^{P}+\Gamma^{NP}\\
&\equiv \frac{2}{d} \sqrt{\frac{\omega}{2\mu_c \sigma_{ends}}}
+\xi \frac{L}{A} \sqrt{\frac{\omega}{2\mu_c \sigma_{cont}}}\,,
\end{split}
\end{equation}
where $\xi$ is defined by:
\begin{equation}
\label{defxi}
\frac{1}{\epsilon\mu\omega^2} \oint_\mathcal{C}\dr \ell\,|\partial_n\psi |^2 \equiv \xi\frac{L}{A}\iint_\mathcal{S}\dr a\,|\psi |^2\,.
\end{equation}
The first term $\Gamma^P$ in equation (\ref{gamma_ohm}) is the \emph{proportional} (homogeneous) damping part of the width (depending only on $\omega$ without any dependence on the spatial structure of the mode) due to effective conductivity $\sigma_{ends}$ of the top and bottom. The second term $\Gamma^{NP}$ is the \emph{non-proportional} part of the width related to a parameter $\xi$, which depends on the spatial structure of the mode at the contour, and therefore varies from mode to mode.

Here it is particularly interesting to revisit equation (\ref{mixedbc}) which turns out to give a quite natural hint of how ohmic losses on the contour induce a small amount of imaginary part for wavefunctions that are purely real in the unperturbed limit. Indeed, assuming an incident plane wave with unit amplitude, the BC (\ref{mixedbc}) lead to a reflected wave with a small (of order $k\delta$)  dephasing. The same BC lead to equation (\ref{eigenpert2}) which relates the width to the quantity $\xi$. 

For the rectangular cavity the values of $\xi$ are readily calculated and display fluctuations around unity, as generally expected\cite{jackson}. Thus we could check the validity of formula (\ref{gamma_ohm}) in our experimental situation\cite{Jerome_2003}. 
In the case of a Sinai-like chaotic cavity  one may view a typical speckle-like mode locally as the superposition of plane waves with random directions, amplitudes and phases. This led  Berry\cite{Berry_1977} to conjecture that the wavefunctions should be \emph{ergodic} on the average, \textit{i.e.} should display all statistical features of a Gaussian random field.  Apart from specific non-ergodic features like \emph{scars} (local intensity enhancement along short unstable periodic orbits), the relative fluctuations of $\xi$ about unity are expected to decrease with the wavelength $\lambda$  as the contour $\mathcal{C}$ encompasses a larger number $\sim$ $L/(\lambda/2)$ of equivalent opening channels.

The cavity used for the measurements presented here is made of OFHC Copper. Ten antennas are fixed through the bottom plate and are used as emitters and receivers as well. These antennas (diameter 1.27\,mm) are only 2\,mm inside the cavity. A rather low coupling is thus achieved, making the role of the antennas nearly negligible regarding their scattering and  loss characteristics. Radiation losses are also made negligible by a tight fitting of the various parts of the cavity. Losses are therefore dominated by ohmic losses on the walls of the cavity. A Vector Network Analyzer HP\,8720\,D is used for the measurements. We restrict our studies to frequencies ranging up to 5.5\,GHz since, at higher frequencies, the modal overlap defined as $\Gamma\times n(\omega)$ ($n(\omega)$ being the modal density per unit interval of $\omega$) becomes too important and invalidates any perturbative approach. Below 5.5\,GHz, apart from few broader resonances, the modal overlap remains smaller than unity. The measurements are performed by 0.5\,GHz-bands, sampled with a frequency step 0.3125\,MHz. The 45 possible couples of antennas are used. On the whole, it amounts to 756\,000 points measured between 0.25\,GHz and 5.5\,GHz.

In order to fit the transmission between two distinct antennas, we use a Breit-Wigner formula  (see e.g. reference\cite{Stoeckbook}) which, in our case, reads\cite{bigone}
\begin{equation}
\label{formule_fit}
S_{ab}(\omega)= \sum_{n} \frac{ \alpha_{n}^{ab} + \ir\beta_{n}^{ab}}{\omega^2-\omega_{n}^2 + \ir \omega_n \Gamma_{n}^{\Omega}}\,.
\end{equation}
The Lorentzian associated to the $n$th resonance for the transmission between the antennas $a$ and $b$ is thus characterized by its complex amplitude $\alpha_{n}^{ab} + \ir \beta_{n}^{ab} =-2 \ir T^2(\omega_n) \psi_n(\vec{r}_a)\psi_n(\vec{r}_b)$, its width $\Gamma_n^{\Omega}$ and its central frequency $\omega_n=\omega_n^0-\Gamma_n^{\Omega}/2$, where $\omega_n^0$ stands for the $n$th eigenfrequency of the unperturbed cavity. $T(\omega)$ is a function representing the coupling of the antennas (all supposed to be identical) to the cavity whose explicit form,  which essentially depends on the geometry of the antenna, has been derived in\cite{Jerome_2003}. We were able to test our model for antenna coupling very precisely  in an empty rectangular cavity where wavefunctions and eigenfrequencies are readily calculated. For the present purpose, it is enough to know that $T$ is a slowly varying function on a frequency scale much larger than the mean spacing $n^{-1}(\omega)$. This coupling is obviously also responsible for additional losses and can be accounted for by introducing corresponding partial widths. The contribution of the latter to the total widths being at most of the order of one percent for all the measurements presented here, we could safely neglect them.

\begin{figure}[h]
\begin{center}
\input{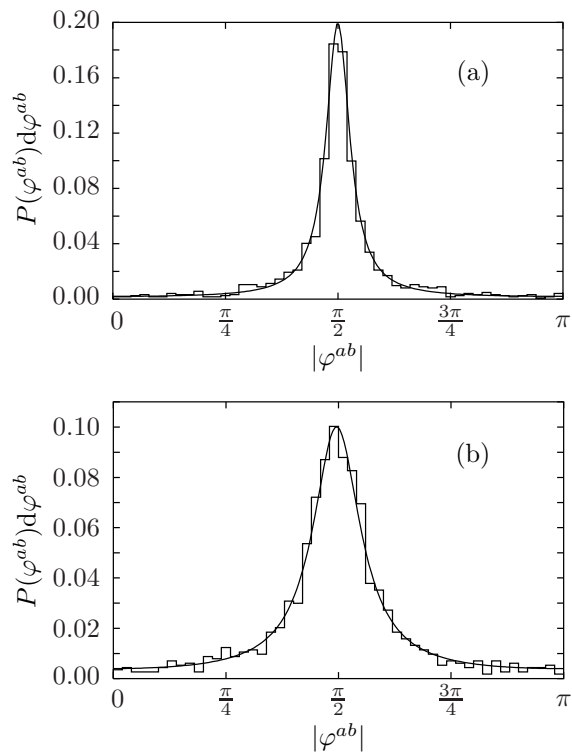}
\end{center}
\caption{Experimental histograms of phase $\varphi^{ab}$ obtained with 30 resonances in the frequency intervals (a) [3.76\,GHz,\,4.1\,GHz] and (b) [5.24\,GHz,\,5.5\,GHz]. Fitted distributions (\ref{distr_phase}) in solid line: (a) $q=5.0\times 10^{-2}$ and (b) $q=10.1\times10^{-2}$.}
\label{qfactor_fig1}
\end{figure}

Assume that the real and imaginary parts of the wavefunction $\psi(\vec{r})$ are Gaussian centered random variables with standard deviations $\sigma_r=\sqrt{\langle (\Re \psi)^2 \rangle}$ and $\sigma_i=\sqrt{\langle (\Im \psi)^2 \rangle}$. The key parameter $q=\sigma_i/\sigma_r$ is a measure of the \emph{openness} of the system. Its value is 0 when the cavity is completely closed (real wavefunctions) and tends to unity for totally absorbing boundaries (complex-valued wavefunctions with equal $\sigma_r$ and $\sigma_i$). In practice, as the spatial sampling over ten antennas for a given resonance is undersized to estimate $\sigma_r$ and $\sigma_i$ accurately, we are led to restrict ourselves to a coarse-grained estimation of $q$ by grouping several contiguous resonances in frequency windows. This may be achieved by building the probability distribution of the phase $\varphi_n^{ab}$ of the complex amplitude of $\alpha_n^{ab} + \ir\beta_n^{ab}$:
\begin{equation}
\label{phase}
\varphi_n^{ab}=-\frac{\pi}{2}+\varphi_n(\vec{r}_a)+\varphi_n(\vec{r}_b)\,,
\end{equation}
where $\varphi_n(\vec{r})$ is the phase of the complex field $\psi_n(\vec{r})$. This distribution is readily obtained from the Gaussian hypothesis and reads\cite{Weaver_2000, Lobkis_2000, Kanzieper_1996}:
\begin{equation}
\label{distr_phase}
P(\varphi^{ab})=\frac{q}{\pi}\frac{1+q^2}{4q^2+(1-q^2)^2 \sin^2(\varphi^{ab}+\frac{\pi}{2})}\,.
\end{equation}
This distribution displays a peak centered on $\vert\varphi^{ab}\vert= \pi/2$ (see figure \ref{qfactor_fig1}). The width of the peak increases with $q$ so that the distribution tends to a Dirac delta as $q\to 0$ or to a uniform distribution as $q\to 1$. The distributions (a) and (b) shown in figure \ref{qfactor_fig1} have been obtained by using 30 resonances lying in the frequency intervals [3.76\,GHz,\,4.1\,GHz] and [5.24\,GHz,\,5.5\,GHz] respectively, yielding coarse-grained estimations of $q$ equal to $5.0\times 10^{-2}$ and $10.1\times10^{-2}$ respectively.

In order to establish the connection between the spatial statistics of complex wavefunctions and their (spectral) widths due to \emph{non-proportional} damping, it is useful to introduce a ray picture of losses due to reflections at the contour. For a plane wave incident on an air-metal interface with angle of incidence $\theta$, it can be shown\cite{Collinbook} that  the amount of absorbed energy reads:
\begin{equation}
\label{power_transmission}
\mathcal{T} = |t|^2 = \frac{4 a\cos{\theta}}{2 a^2\cos^2{\theta}+2 a \cos{\theta}+1}\,,
\end{equation}
where $a = \sqrt{\omega \epsilon/(2 \sigma_{cont})}$. In the case of OFHC copper $a\sim 10^{-5}$ around 1\,GHz and the previous equation reduces to 
\begin{equation}
\label{trans_moy}
\mathcal{T} \simeq 4 a \cos{\theta}\,.
\end{equation}
For an ergodic wavefunction in a chaotic billiard, the appropriate statistical measure for angular average
is $\dr(\sin{\theta)}/2$ leading to a mean absorption coefficient
\begin{equation}
\label{trans_moy2}
\langle\mathcal{T}\rangle =\pi\sqrt{\omega \epsilon/(2\sigma_{cont})}\,.
\end{equation}
Having in mind the analogy of our problem with ultrasonics in a reverberant body\cite{Lobkis_00jsv} or room acoustics\cite{Mortessagne_1993}, one may use the so-called Sabine's law\cite{Mortessagne_1993} to deduce the average width $\Gamma_{refl}$ related to absorption at the contour
\begin{equation}
\label{gamma_reflexion} 
\Gamma_{refl} = \frac{cL}{\pi A}\langle\mathcal{T}\rangle = \frac{L}{A}\sqrt{\frac{\omega}{2\mu\sigma_{cont}}}\,.
\end{equation}
This last result may be identified with the expression of $\Gamma^{NP}$ in equation (\ref{gamma_ohm}) provided that $\xi$ be replaced by its mean value for an  ergodic mode namely unity\cite{bigone}. As $q$ and the transmission $\mathcal{T}$ both vary within the same interval ranging from 0 (completely closed system) to 1 (completely transparent contour), we propose to identify $q$ and $\mathcal{T}$, leading to the following relationship:
\begin{equation}
\label{gamma_refl_moy2}
\Gamma^{NP} = \frac{c L}{\pi A } q\,.
\end{equation}
This simple identification is all the more natural that the spectral perturbation $\Gamma^{NP}$ and the spatial perturbation $q$ are of the same order. Indeed,  the relationship (\ref{mixedbc}) indicates that $q$ is of order $k \delta$. Likewise, equation (\ref{gamma_ohm}) shows that $(A/cL)\,\Gamma^{NP}$ is also of order $k \delta$.  This excludes any power law relationship, other than proportionality, between $q$ and $\Gamma^{NP}$. 

\begin{figure}[h]
\begin{center}
\input{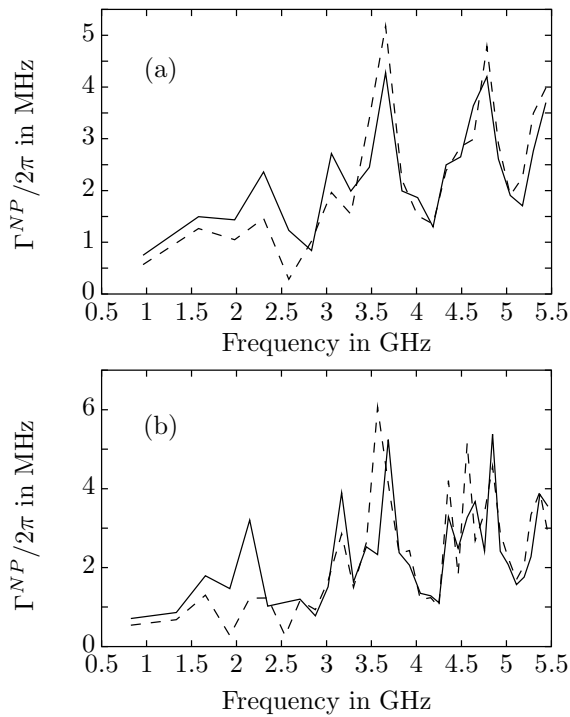}
\end{center}
\caption{Solid lines correspond to evaluations of $\Gamma^{NP}$ extracted from our measurements through expression (\ref{formule_fit}) with an average over contiguous groups of: (a) 15 resonances and  (b) 10 resonances. Dashed lines shows relation  (\ref{gamma_refl_moy2}) with coarse-grained $q$ deduced from the distribution (\ref{distr_phase}).}
\label{qfactor_fig2}
\end{figure}

From our measurements, we used formulas (\ref{formule_fit}) and (\ref{gamma_ohm}) to extract, by an appropriate and original fitting procedure\cite{Jerome_2003}, the values of the \emph{non-proportional} widths $\Gamma^{NP}$ for all 330 resonances up to 5.5\,GHz. These data are shown on figure \ref{qfactor_fig2}, where an average over contiguous groups of 15 (a) and 10 (b) resonances is performed. They are compared to expression (\ref{gamma_refl_moy2}) where a coarse-grained evaluation of $q$ is deduced from the distribution (\ref{distr_phase}) built with the same groups of resonances. With only 10 resonances, we have checked that fitting the experimental histograms of phase with formula (\ref{distr_phase}) is still legitimate. The obvious correlation between the effective measure of imaginary parts, given by the parameter $q$, and the variations of the widths gives strong support to our hypothesis embodied in relation (\ref{gamma_refl_moy2}). The agreement is fairly good even though some small discrepancies can be observed, which can be attributed to the lack of ergodicity of certain modes. Our belief is that the relation (\ref{gamma_refl_moy2}) should hold for each truly \emph{ergodic} mode but its verification is out of reach with our present experimental set.

That the wavefunctions of a lossy wave system are in general complex has not been well appreciated in the past literature maybe because the most common and simple model of losses leaves the modes real. Such a model predicts that the resonance widths should only vary slowly with frequency. In this letter, for the first time to our knowledge, we propose theoretical arguments in favor of the hypothesis that the imaginary parts of the modes should correlate with the \emph{non-proportional} part of the widths. Through an original analysis of transmission measurements in a lossy chaotic microwave cavity, we provide clear experimental evidence of its pertinency. It should be stressed that our point of view is complementary of RMT in that it focuses on spatial aspects whereas RMT for open systems is mainly concerned with spectral statistics\cite{Dittes00}. We believe that the present work may have conceptual implications in transport in open mesoscopic devices\cite{Alhassid_2000} and also in structural acoustics\cite{Lobkis_2003}.

%\acknowledgments{The authors gratefully acknowledge helpful discussions
%with P. Sebbah, R. L. Weaver and H.-J. St{\"o}ckmann and wish to thank V. Doya and Ch. Vanneste for useful comments about the manuscript.}

\bibliography{qfactor}

\end{document}